\documentclass[12pt]{article}
\usepackage{graphicx}
\usepackage{feynmp}
\usepackage{amsmath}
\usepackage{mathrsfs}
\def\beq{\begin{equation}}
\def\eeq{\end{equation}}
\def\bea{\begin{eqnarray}}
\def\eea{\end{eqnarray}}

\def\roughly#1{\mathrel{\raise.3ex\hbox
{$#1$\kern-.75em\lower1ex\hbox{$\sim$}}}}
\def\lsim{\roughly<}

\def\sss{\scriptscriptstyle}

\def\bra#1{\left\langle  #1\right|} \def\ket#1{\left| #1\right\rangle}
\def\barpk{{\raise.35ex\hbox  {${\sss  (}$}}--{\raise.35ex\hbox{${\sss
)}$}}}        \def\bbarp{\hbox{$B$\kern-0.9em\raise1.4ex\hbox{\barpk}}}

\def\lsim{\roughly<}

\def\bra#1{\left\langle  #1\right|} \def\ket#1{\left| #1\right\rangle}

  \def\rr2{{1\over\sqrt{2}}}

\def\.{\!\cdot\!}    \def\:{\cdots}   \def\[{\left[}   \def\]{\right]}
\def\({\left(} \def\){\right)} 
\newcommand{\etatautau}{\eta_b \to \tau^+\tau^-}
\newcommand{\uptautaugamma}{\Upsilon \to \tau^+\tau^- \gamma}
\newcommand{\upgammaU}{\Upsilon \to A^0(U) \gamma}
\newcommand{\adecay}{A^0(U) \to \tau^+ \tau^-}
\newcommand{\au}{A^0(U)}

\begin{document}
 \unitlength = 1mm
\begin{flushright}
UMISS-HEP-2010-03 \\
[10mm]
\end{flushright}

\begin{center}
\bigskip {\Large  \bf Probing light pseudoscalar, axial vector states through $\eta_{b} \rightarrow\tau^{+}\tau^{-}$.}
\\[8mm]
Ahmed Rashed $^{\dag, \ddag}$
\footnote{E-mail:
\texttt{amrashed@phy.olemiss.edu}}
, Murugeswaran Duraisamy $^{\dag}$ 
\footnote{E-mail:
\texttt{duraism@phy.olemiss.edu}} 
, and Alakabha Datta $^{\dag}$ 
\footnote{E-mail:
\texttt{datta@phy.olemiss.edu}} 
\\[3mm]
\end{center}

\begin{center}
~~~~~~~~~~~ {\it $^{\dag}$ Department  of Physics and Astronomy,}\\ 
~~~~~~~~~~~~{ \it University of Mississippi,}\\
~~~~~~~~~~~~{\it  Lewis Hall, University, MS, 38677.}\\
\end{center}

\begin{center}
~~~~~~~~~~~ {\it $^{\ddag}$ Department  of Physics,}\\ 
~~~~~~~~~~~~{ \it Ain Shams University,}\\
~~~~~~~~~~~~{\it  Faculty of Science, Cairo, 11566, Egypt.}\\
\end{center}


\begin{center} 
\bigskip (\today) \vskip0.5cm {\Large Abstract\\} \vskip3truemm
\parbox[t]{\textwidth}  {
In this paper we explore the decay $\etatautau$ as a probe for a light pseudoscalar or a light axial vector state. 
We estimate the standard model branching ratio for this decay to be $\sim 4 \times 10^{-9}$. We show that considerably 
larger branching ratios, up to the present experimental limit of $\sim 8$\%, is possible in models with a light pseudoscalar 
or a light axial vector state. As we do not include possible mixing effects between the light pseudoscalar and the $\eta_b$, 
our results should be reliable when the pseudoscalar mass is away from the $\eta_b$ mass. 
} 
\end{center}

\thispagestyle{empty} \newpage \setcounter{page}{1}
\baselineskip=14pt

\section{Introduction}
It is widely anticipated that physics beyond the standard model (SM) or new physics (NP) will be discovered soon at experiments 
such as the LHC.
This NP might contain new gauge bosons, additional Higgs bosons beyond the SM Higgs, or new quarks and leptons. It is generally 
believed that these new particles will be heavy with masses from the weak scale $\sim 100$ GeV to a TeV. However, light scalars and 
vector bosons with masses in the GeV range or even lower are not ruled out. For instance, light scalar states coming from a primary higgs with non SM decays can be consistent with existing experimental constraints \cite{Chang}.
 One of the ways to probe these light states is to look 
at decays of particles with masses in the 10 GeV range such as the $\Upsilon$. Data from the present and future $B$ factories can 
be used to search for these states and/or to put constraints on models that predict such states.

The  pseudoscalar $b \bar{b}$ bound state in the 1S configuration, the $\eta_b$, was recently observed.
Two research groups in BaBar observed it
in two different experiments. First, it was seen in the
decay of $\Upsilon(3S)\to\gamma\eta_{b}$ \cite{:2008vj}
with a signal significance
greater than 10 standard deviations $(\sigma)$. The $\eta_b$ was observed  in the photon energy spectrum using 
$(109\pm1)$ million $\Upsilon(3S)$ events and the hyperfine
$\Upsilon(1S)-\eta_{b}$ mass splitting was measured to be
$71.4^{+2.3}_{-3.1}\text{(stat)}\pm2.7\text{(syst)}$ MeV from the mass $m(\eta_b)={9388.9}^{+ 3.1}_{-2.3}\.({\rm stat}) \pm 2.7\,{(\rm syst)} ~{\rm MeV}$. Soon after, it was also seen in
$\Upsilon(2S)\to\gamma\eta_{b}$ \cite{:2009pz} by another group in
BaBar, and the hyperfine mass splitting was determined to be
$67.4^{+4.8}_{-4.6}(\text{stat})\pm2.0(\text{syst})$ MeV from the mass  $m(\eta_b)={9392.9}^{+ 4.6}_{-4.8}\,
({\rm stat})\pm 1.9\,{(\rm syst)} ~{\rm MeV}$ .
In the past, since the discovery of the $\Upsilon(nS)$ resonances \cite{1977} in 1977, various experimental environments \cite{Mahmood:2002jd, Heister:2002if, Tseng:2003md} have been used to seek the
ground state $\eta_{b}$ but without success. 
 Many theoretical models have attempted to predict the mass of $\eta_{b}$. 
Lattice NRQCD \cite{Liao:2001yh, Liao:2001yh2}  predicts the hyperfine splitting to be $E^{lat}_{hfs}=61\pm 14$ MeV and correspondingly the mass to be $m_{\eta_b}=9383(4)(2)$ MeV which is in agreement with the experimental results. The calculations of
perturbative QCD based models \cite{Liao:2001yh2, Kniehl:2003ap} predict the hyperfine splitting to be $E^{QCD}_{hfs}=39\pm 11(\text{th})^{+9}_{-8} (\delta \alpha_{s})$ MeV which is smaller than the measured values.
Experiments at BaBar have also searched for a low-mass Higgs boson in
$\Upsilon(3S)\rightarrow\gamma A^0$, $A^0\rightarrow\tau^+\tau^-$ 
\cite{decay} with data sample containing 122 million
$\Upsilon(3S)$ events. In the same analysis,  constraint on the branching ratio for 
 $\eta_b\rightarrow \tau^+\tau^-$ 
was reported as ${\mathcal{BR}}(\eta_b\rightarrow \tau^+\tau^-)<8\%$ at $90\%$ confidence level (C.L.).

In this paper we will be interested in probing  light scalar and spin 1 states via $\eta_b$ decays. As the $\eta_b$ is a pseudoscalar a light pseudoscalar and a spin 1 state with axial vector coupling can directly couple to $\eta_b$.
We will assume the pseudoscalar to couple to the mass of the fermion as is usually the case for Higgs coupling to fermions. Hence, the $\eta_b$ which is a $b \bar{b}$ bound state has  advantages over the $\eta_c$ and $\eta/{\eta'}$ mesons which are $c\bar{c}$ and $q\bar{q}(q=u,d,s)$ bound states, respectively. The $\eta_b$ is expected to be a  sensitive probe of a light axial vector state. This follows from the fact that the longitudinal polarization of the axial vector, $\epsilon^{\mu}_L\sim k^{\mu}$, when $k^{\mu}$ the  momentum of the vector boson is much larger than its mass. Consequently, the effective axial vector-fermion pair coupling is proportional to the fermion mass for the longitudinal polarization.

 In this work we will study the process $\eta_b \to \tau^+ \tau^-$ mediated by a pseudoscalar ($A^0$) or an axial vector ($ U $). In the SM this process can only go through a $Z$ exchange at tree level and is highly suppressed with a branching ratio $\sim 4 \times 10^{-9}$. There is also
 a higher order contribution to $\etatautau$ in the SM, via two intermediate photons. The branching ratio for this process is also tiny $\sim 10^{-10}$. Hence, a measurement of $BR[\etatautau]$  larger than the SM rate will be a signal of new states. One can also probe the states $\au$ in $\Upsilon$ decays. To search for light $ \au$ states in $\Upsilon$ decays one generally considers the decay chains, $\upgammaU \; (\adecay)$
 \cite{decay}. In other words, the $\au$ is assumed to be produced on-shell. One then looks for a peak in the
invariant mass of the $\tau$ pairs.
The experimental measurement/constraint of 
$BR[\upgammaU] \times BR[\adecay]$
 can  be converted into a measurement/constraint on the coupling of the $\au $ to $b \bar{b}$, and hence on model parameters, if the $BR[\au \to \tau^+ \tau^-]$ is used as an input \cite{gun1}. 
  Clearly as $m_{\au} > m_{\Upsilon} $, the $\au$  can no longer be produced on-shell and the rate for
$\uptautaugamma$ will fall  and consequently the constraints on the model parameters will be weaker.
Note that the constraint $m_{A^0} < 2 m_B$ needs to be assumed in the very particular case where
  the $CP$-even Higgs mass $m_h<114$ GeV and $h\rightarrow 2A^0$ dominates over $h\rightarrow 2m_b$ \cite{Chang}. In general
 $m_{A^0} > 2 m_B$ is also possible.
 We will just assume the existence of light pseudoscalar and axial vector states close to the $\eta_b$ mass but they can have masses that are  greater than or less than $ 2 m_b$.
  
The $\eta_b$ has only been seen in the radiative decays $\Upsilon \to \gamma \eta_b$. Hence, the decay $\etatautau$ has only been studied via the decay $\uptautaugamma$. However, the decay $\etatautau$ can be studied independently from the process $\uptautaugamma$ as the $\eta_{b}$ can be produced from various other processes such as two-photon collisions, $\gamma\gamma\rightarrow\eta_{b}$ \cite{Heister:2002if}, and in two parton collisions \cite{Tseng:2003md, Braaten:2000cm}, in hadron colliders like the Tevatron and  the LHC.
The process $\etatautau$ has several advantages over $\Upsilon$ decays in probing $\au$ states specially when  $\au$ is off-shell which is always the case when $m_{\au}> m_{\Upsilon}$.
First, unlike the $\eta_b$ which can couple directly to $\au$, the $\Upsilon $ can  only couple to $\au$ in conjunction with another state- usually a photon.
Hence, the $\Upsilon $ couplings are second order and therefore
 it can decay only to the $\tau^+\tau^- \gamma$ state with a rate  much smaller than the rate for $\etatautau$. However,
 the $\Upsilon$ states are narrower than the $\eta_b$, which may compensate partially the larger rate for $\eta_b \to \tau^+ \tau^-$ relative to
$\Upsilon \to \tau^+ \tau^- \gamma$ in the branching ratio measurements.
Secondly, an important distinction between $\uptautaugamma$ and $\etatautau$ is that the former decay can also proceed as a radiative decay in the SM while the latter decay is highly suppressed in the SM as indicated above.
Adapting the expression used to estimate the SM branching ratio for
$J/\psi \rightarrow e^+e^- \gamma$ \cite{jpsi},  with the $\gamma$ emitted from the final state electrons, to the decay $\uptautaugamma$, the rate for this decay in the SM is,
\bea
d\Gamma_{\Upsilon\rightarrow \tau^{+}\tau^{-}\gamma} & = & d\Gamma_{\Upsilon\rightarrow \tau^{+}\tau^{-}}\beta'^{3}\frac{2 \alpha}{\pi}\frac{dE'_{\gamma}}{E'_{\gamma}}\frac{s'}{s} \frac{1-cos^{2}\theta'_{\gamma \tau}}{(1-\beta'^{2}cos^{2}\theta'_{\gamma \tau})^{2}}d\Omega'_{\gamma}\
\label{into1}
\eea
with
\bea
 d\Gamma_{\Upsilon\rightarrow \tau^{+}\tau^{-}} & = & \frac{3}{3+\lambda}(1+\lambda\cos^{2}\theta'_{\tau})\Gamma_{\Upsilon\rightarrow \tau^{+}\tau^{-}}\frac{d\Omega'_{\tau}}{4\pi}.\
\label{into2}
\eea
Here $E'_{\gamma}$ represents the $\gamma$ energy, $\theta'_{\gamma}$ and $\phi'_{\gamma} (\Omega'_{\gamma})$ the $\gamma$ angles, and  $\theta'_\tau$ and $\phi'_\tau (\Omega'_{\tau})$ the $\tau$ angles, all in the $\tau^{+}\tau^{-}$ c.m. frame. $\beta'$ is the $\tau$ velocity and $\theta'_{\gamma \tau}$ is the angle between the $\tau$ and $\gamma$ directions, also in the $\tau^{+}\tau^{-}$ c.m. frame while $s'$ is the $\tau^{+}\tau^{-}$ invariant mass squared and $s$ is the $\Upsilon$ invariant mass squared. The parameter $\lambda$ is determined from the experimental data to be $(0.88\pm 0.19)$ \cite{jpsi}.
Using the branching ratio for ${\Upsilon\rightarrow \tau^{+}\tau^{-}}=2.6\times 10^{-2}$ \cite{pdg} we estimate the branching ratio for ${\Upsilon\rightarrow \tau^{+}\tau^{-}\gamma}=4.4\times10^{-3}$ 
with $E_{\gamma}> 100$ MeV.

Naively, the rate for $\uptautaugamma$ through an off-shell $A^0$, from a 2HDM of type II, relative to the SM rate for $\uptautaugamma$ is $\sim$ ${{g^4 \tan^4 \beta m_b^2 m^{2}_{\tau}}} \over {16e^4 M_W^4} $. Therefore, for large $\tan \beta \sim 28 $ the SM and the NP rates may be comparable. However given the hadronic uncertainties in estimating
the SM and the NP rates for $\uptautaugamma$, it will be difficult to distinguish between the NP and the SM contributions.
Hence, searching for $\au$ with $m_{\au}> m_{\Upsilon}$ in $\uptautaugamma$ will be very difficult because of the large SM background.
Note that even in $e^{+}e^{-}$ machines like the $B$-factories where the $\eta_{b}$ is produced through the decay $\Upsilon\rightarrow\gamma\eta_{b}$, the product of branching ratios  $BR[\Upsilon\rightarrow\gamma\eta_{b}]\times[\eta_{b}\rightarrow\tau^{+}\tau^{-}]$ is tiny in the SM because of the highly suppressed $BR[\eta_{b}\rightarrow\tau^{+}\tau^{-}] \sim 4 \times 10^{-9}$.
Using the measured $BR[\Upsilon\rightarrow\gamma\eta_{b}] \sim 5 \times 10^{-4}$ \cite{:2008vj,:2009pz} one obtains
$BR[\Upsilon\rightarrow\gamma\eta_{b}]\times[\eta_{b}\rightarrow\tau^{+}\tau^{-}] \sim 2 \times 10^{-12} $ which is very difficult to measure.
In the presence of new physics
this product of branching ratios is enhanced and  can reach $\lsim 10^{-5}$. Hence the observation  of $\Upsilon\rightarrow\gamma\tau^{+}\tau^{-}$, with the $\tau$ pairs coming from $\eta_b$, at branching ratios much larger than the SM expectations will be signal for new light states. In summary, the large SM background in $\uptautaugamma$ and a tiny SM contribution to $\etatautau $ makes the later decay   potentially a better probe for $\au$ than the former if 
the decays proceed through the off-shell exchange of $\au$.

There are good theoretical motivations for the  existence of a light CP-odd  $A^0$ Higgs boson or an axial vector boson $ U $ with masses,
$m_{A^0}$ and $m_{U}$ respectively, in the GeV range or below.
There has been
interest in the $m_{A^{0}}<2m_B$ region, for which a light Higgs, $h$, with
SM-like $WW$, $ZZ$ and fermionic couplings can have mass $m_h \sim
100$ GeV while still being consistent with LEP data by virtue
of $h \to A^0 A^0 $.
This scenario could even explain the $2.3\, \sigma$ excess in
the $e^+ e^- \to Z + 2b$ channel for $M_{2b} \sim
100$~GeV \cite{lep}.  Such a light pseudoscalar Higgs can naturally arise in extensions of MSSM with additional singlet
scalars and fermions (gauge-singlet supermultiplets) known as Next-to-Minimal Supersymmetric Model (NMSSM) \cite{nmssm}.
Constraints on models with a light $A^0$ state have been studied recently within a 2HDM framework with certain assumptions about the coupling and in NMSSM \cite{gun1,Gunion:2008dg,Domingo:2008rr}.  

Our goal will not be to work in a specific model but we will assume the couplings of the $A^0$ to the $b$ quark and the $\tau$ lepton to be the same as in the 2HDM. We will assume this 2HDM is part of some extension of the SM. Hence, we will not strictly follow the bounds and constraints
obtained in some specific extension of the SM which includes the 2HDM,  but will choose values for the parameters in our calculation which are similar to constraints on these parameters in specific NP models. 
The process $\etatautau$ will proceed through an off-shell $A^0$ and we will consider both $m_{A^0} < m_{\eta_b}$
and $m_{A^0} > m_{\eta_b}$. In general, there will be mixing between $A^0$ and the $\eta_b$ and as 
 the pseudoscalar state gets close to the $\eta_b$ mass  the mixing between the states will become important \cite{mix}. The calculation of this mixing is model dependent and while there are estimates of this mixing in simple quark models the mixing may be very different in other approaches to the bound state problem in QCD. Hence, we will not take into account mixing in our analysis.
   Therefore, our results will be reliable when the $A^0$ mass is away from the $\eta_b$ mass.
We will further assume that the $A^0$ is narrow and neglect its width in our calculations. This approximation will be good as long as $m_{A^0}$ is sufficiently away from the $\eta_b$ mass. When $A^0$ is produced on-shell both mixing and width effects will become important and our results will not be  reliable.

There are also models, for example within SUSY with extra gauged $U(1)$, which have  a light axial vector state \cite{ussm}. These light states can also mediate the process $\eta_b \to \tau^+ \tau^-$. Constraints on these models have been studied \cite{fayet1, Bouchiat,fayet2,fayet3,fayet4}. We will consider $\etatautau$ through the exchange of the axial vector $U$. To perform our calculations we will choose the model discussed in \cite{Bouchiat, fayet4} and neglect the width of the $U$-boson.

Finally, we note that there are recent dark matter models \cite{darkmatter} that also contain light scalar (pseudoscalar) and vector (axial vector) states which may be probed via $\etatautau$.
The HyperCP collaboration has some events 
 for the decay $\Sigma^+ \to p \mu^+ \mu^- $ which may be interpreted
 as evidence for a light pseudoscalar state \cite{hypercp}.

This paper is organized in the following manner. In section 2 we perform the calculations 
of the decay $\eta_{b}\rightarrow \tau^{+}\tau^{-}$ in the SM and in models with a light pseudoscalar $A^0$ and a light axial vector $U$ state. In section 3 we present the numerical results 
of the branching ratios for $\etatautau $. Finally, in section 4 we present our conclusion.

\section{$\etatautau$ in the SM and NP} 
In this section we will study $\etatautau$  in the SM and in models of NP. The $\eta_b$ is a pseudoscalar and cannot couple to $\gamma$ directly. Hence, in the SM, $\etatautau$ can only proceed through the exchange of a $Z$ at tree level and we will calculate the branching ratio for this process in the SM . This decay can also proceed at higher order in the SM through  intermediate two photon states. 

\begin{figure}[h!]
\centering
\includegraphics[width=12cm]{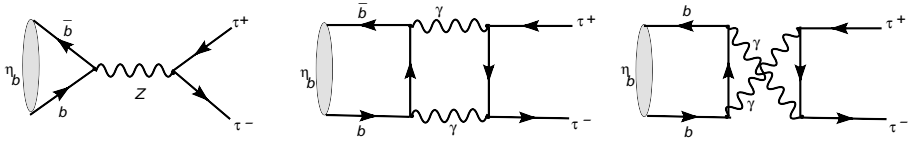}
%
\caption{Various processes contributing to $\eta_{b}\rightarrow \tau^{+}\tau^{-}$ in the SM.}
\label{etabdia}
\end{figure}

\begin{figure}[h!]
 \centering
\includegraphics[width=7cm]{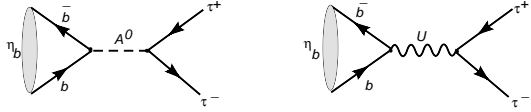}
%
\caption{Various processes contributing to $\eta_{b}\rightarrow \tau^{+}\tau^{-}$ in NP.}
\label{etabdiaNP}
\end{figure}

In the presence of NP $\etatautau$ can proceed through the exchange of a light pseudoscalar or a light spin 1 boson with axial vector coupling. We will consider these two NP scenarios in this section.
The various tree level contribution to the $\eta_b\rightarrow \tau^+\tau^-$ in the SM and NP are shown in Fig. \ref{etabdia} and Fig. \ref{etabdiaNP}, respectively.

We begin with $\etatautau$ in the SM.
We show, in  Fig.~\ref{etabdia}, the decay process $\eta_b\rightarrow \tau^+\tau^-$  via the Z-boson exchange and through the two photon intermediate states. The  decay rate for the tree level $Z$ exchange process can be obtained as,
\bea
\label{DRZ} 
\Gamma^{Z}(\eta_b\rightarrow \tau^+\tau^-)&=& \frac{G^2_F M^4_W m^2_\tau f^2_{\eta_b} m_{\eta_b}}{16 \pi \cos^4{\theta_W} }\beta_\tau  \Big(1-\frac{m^2_{\eta_b}}{M^2_Z}\Big)^2|a_{Z}|^2,\
\eea
where $\theta_W$ denotes the Weinberg angle, 
$\beta_\tau=\sqrt{1-\left( \frac{2 m_\tau}{m_{\eta_b}}\right)^{2}}$ is the velocity of the $\tau$ lepton in the $\eta_b$ rest frame and 
\bea
\label{aZ}
|a_{Z}|^2 &\equiv& \frac{1}{(m_{\eta_b}^2-M^2_{Z})^2+ M^2_{Z} \Gamma^2_{Z}}.
\eea
The decay constant $f_{\eta_b}$ in Eq.~\ref{DRZ} is defined as \cite{Fayet:2009tv-2},
\bea
\label{feta}
\bra {0}\bar{b}(0)\gamma_\mu \gamma_5 b(0)\ket{\eta_b(q)} &=& i f_{\eta_b} q_\mu.\
\eea 

The process $\etatautau$ can also go via two photon intermediate states as shown in diagram Fig. \ref{etabdia}. This diagram is dominated by the imaginary part \cite{YuJia} which  we can estimate  using unitarity \cite{uni} to obtain,
 \bea
 \Gamma^{2 \gamma}[ \etatautau]  & \ge &
 \frac{\alpha^2}{2 \beta_{\tau}} \left[ \frac{ m_{\tau}}{m_{\eta_b}} 
 \ln\frac{(1+ \beta_{\tau})}{(1-\beta_{\tau})} \right]^2 \Gamma[\eta_b \to \gamma \gamma], \
 \label{etahigh}
 \eea 
 where $\alpha$ is the electromagnetic fine structure constant.
 One can calculate $\Gamma[\eta_b \to \gamma \gamma]$ as,
 \bea
 \Gamma[\eta_b \to \gamma \gamma] & = & 
 \frac{\pi \alpha^2 m_{\eta_b}f_{\eta_b}^2}{81m_b^2}, \
 \label{etagamma}
 \eea
where we have used the heavy quark limit for the $b$ quark. 
Since the 2$\gamma$ exchange contribution is mostly imaginary relative to the $Z$ exchange contribution therefore to a good approximation the total width $\Gamma_t[\etatautau]$ is ,
\bea
\Gamma_t[\etatautau]  & \approx & \Gamma^Z[\etatautau]+ \Gamma^{2\gamma}[\etatautau]. \
\label{gammatotal}
\eea
We now turn to NP models and begin with the 2HDM.
The couplings of the down-type quarks D and charged leptons $\ell$  with  ${A^{\text{0}}}$ in the generic
2HDM model are given by \cite{Diaz}
\beq
 \label{2HDMcoup}
\mathcal{L}^{D,\ell}_{A^{\text{0}}}=\frac{i g F_{A^{\text{0}}}}{2 M_W}(\bar{D} M^{diag}_D \gamma_5 D+ \bar{\ell} M^{diag}_l \gamma_5 \ell) A^{0},
\eeq
where $F_{A^{0}}$ is a model-dependent parameter,
  $ M^{diag}_D=(m_d, m_c,m_b)$ and $ M^{diag}_{\ell}=(m_e,m_\mu,m_\tau)$ are the diagonal mass matrices of  D and $\ell$, respectively.
 We will consider 
  $F_{A^{0}} > 1$ in our analysis.
In the case of 2HDM type (II) $F_{A^{0}}\equiv tan \beta$ while in 
2HDM type (I) $F_{A^{0}}\equiv -\cot \beta$.    
  
  In   Fig.~\ref{etabdiaNP}(a) we show the decay process $\eta_b\rightarrow \tau^+\tau^-$  via the exchange of 
the $CP$-odd Higgs scalar $A^{\text{0}}$.
  The decay rate for this process can be obtained as,
\bea
\label{DR2HDM} 
\Gamma^{A^0}(\eta_b\rightarrow \tau^+\tau^-)&=& \frac{G^2_F m^2_\tau f^2_{\eta_b} m^5_{\eta_b}}{16 \pi}\beta_\tau |a_{A^0}|^2,
\eea
where the coefficient $a_{A^0}$ depends on the mass $m_{A^0}$ as, 
\bea
\label{aiIII}
|a_{A^0}|^2 &\equiv& \frac{F_{A^{0}}^4}{(m_{\eta_b}^2-m^2_{A^0})^2}.
\eea
 We have assumed that the decay width $\Gamma_{A^0}$ for the $A^0$ is negligible. In Eq.~\ref{DR2HDM},
we have used,
\bea
\label{feta2}
\bra {0}\bar{b}(0) \gamma_5 b(0)\ket{\eta_b(q)} &=& \frac{i f_{\eta_b} m^2_{\eta_b}}{2 m_b} ,
\eea
where $f_{\eta_b}$ has been defined in Eq.~\ref{feta}.

Finally, we move to NP models that contain a light spin 1 boson with axial vector couplings. 
In Fig. \ref{etabdiaNP}(b) we show the  decay process $\eta_b\rightarrow \tau^+\tau^-$  via the exchange of the  light neutral gauge boson $U$.
We  write down a model independent Lagrangian for the $U$-boson but we assume the structure of the Lagrangian to be similar to the one discussed in Ref.~\cite{Bouchiat, fayet2,fayet3}.
 We take the $U$ couplings to the down-type quarks and charged leptons to be given by  
\bea
\label{Ucoup}
\mathcal{L}^{D,\ell}_{U} &=&  f^{D,\ell}_{A}(\bar{D}\gamma^\mu \gamma_5 D+ \bar{\ell}\gamma^\mu \gamma_5 \ell)U_{\mu},
\eea
with the axial coupling 
\bea
\label{ucoup}
f^{D,\ell}_{A} &=& 2^{-\frac{3}{4}} G^{\frac{1}{2}}_F m_U F_{U},
\eea
where $m_U$ denotes the mass of $U$-boson and $F_{U}$ denotes a model-dependent parameter. 
In the specific model \cite{Bouchiat, fayet2,fayet3}, $F_{U}\equiv \cos{\zeta} \tan\beta$.

Again, we will be interested in $F_{U} > 1$.
The decay rate for $\etatautau$  can be obtained as
\bea
\label{DRUI} 
\Gamma^{U}(\eta_b\rightarrow \tau^+\tau^-)&=& 
 \frac{ G^2_F m^2_\tau f^2_{\eta_b} m_{\eta_b}}{16 \pi }\beta_\tau  (m^{2}_{U}-m^2_{\eta_b})^2 F_{U}^4|a_{U}|^2,
\eea
where 
\bea
\label{aZ1}
|a_{U}|^2 &=& \frac{1}{(m_{\eta_b}^2-m^2_{U})^2+ m^2_{U} \Gamma^2_{U}}.
\eea
Eq.~\ref{aZ1} can be expanded as,
\bea
\label{aZ2}
|a_{U}|^2 &=& \frac{1}{(m_{\eta_b}^2-m^2_{U})^2}(1-x^2+\dots),
\eea
if $x= \frac{\Gamma_{U}/m_{U}}{(1-m^2_{\eta_b}/m_{U}^2)} < 1$.

 Neglecting $x$, Eq.~\ref{DRUI} reduces to
\bea
\label{DRUII} 
\Gamma^{U}(\eta_b\rightarrow \tau^+\tau^-)&=&\frac{ G^2_F m^2_\tau f^2_{\eta_b} m_{\eta_b}}{16 \pi }\beta_\tau  F_{U}^4.
\eea 
Thus, Eq.~\ref{DRUII} shows that the decay width for $\etatautau$  does not depend on $m_{U}$ in the approximation of neglecting the width of the $U$-boson. This result is easy to understand.
If one increases the mass of the $U$ then the matrix element for $\etatautau$ is suppressed due to propagator effects. However, the coupling, which is proportional to $m_U$, increases  to compensate for this suppression. The fact that the width for $\etatautau$ is independent of $m_U$ only holds because the $\eta_b$ is a pseudoscalar. 

The result of Eq.~\ref{DRUII} does not make sense as $m_U$ gets sufficiently large as the couplings in Eq.~\ref{ucoup} becomes non-perturbative. Requiring the couplings to be $ \le 1 $ one gets the constraints 
$ m_U \le { {4 M_W} \over {g F_{U}}}$. Hence for $F_{U} \sim 50$ one can get $m_U$ to be in the GeV range.

 It is interesting to note that  in the up sector the behavior for the decay width is different. The coupling of the vector 
boson to the up type quark, $U$, is given by
\bea
\label{Ucoup2}
\mathcal{L}_{U} &=&  f^{UP}_{A}\bar{U}\gamma^\mu \gamma_5 U U_{\mu},
\eea
with the axial coupling of the up-type quarks
\bea
\label{ucoup2}
f^{UP}_{A} &=& 2^{-\frac{3}{4}} G^{\frac{1}{2}}_F m_U F'_{U}.
\eea 
 In the model of Ref.~\cite{Bouchiat, fayet2,fayet3}, $F'_{U}\equiv \cos{\zeta} \cot\beta$.

For instance, the branching ratio $\mathcal{BR}(\eta_{c}\rightarrow \mu^{+}\mu^{-})$ does not depend on $m_{U}$ or on $tan \beta$ and is given as,
\bea
\label{DRUII-2} 
\Gamma^{U}(\eta_c\rightarrow \mu^+\mu^-)&=&\frac{ G^2_F m^2_\mu f^2_{\eta_c} m_{\eta_c}}{16 \pi }\bar{\beta}_{\tau}   \cos^4{\zeta}.
\eea
where $\bar{\beta}_{\tau}  =\sqrt{1-\left( \frac{2 m_\mu}{m_{\eta_c}}\right) ^{2}}$ and $f_{\eta_{c}}$ is the $\eta_c$ decay constant. We can see from Eq.~\ref{DRUII-2} that the branching ratio $\mathcal{BR}(\eta_{c}\rightarrow \mu^{+}\mu^{-})$ 
is much smaller than $\mathcal{BR}(\eta_{b}\rightarrow \tau^{+}\tau^{-})$ if $tan \beta >1$ because of the absence of the factor $\tan^{4} \beta$ in the rate for $\eta_{c}\rightarrow \mu^{+}\mu^{-}$ .

\section{Numerical Analysis}
In this section we present our numerical results.
We take the average $\eta_b(1S)$ mass to be $m_{\eta_b}=9390.8 \pm 3.2 $ MeV \cite{:2009pz}, the
decay constant $f_{\eta_{b}}=(705\pm 27)$ MeV \cite{decay-const} and the width to be   $\Gamma_{\eta_b}\approx 10$ MeV \cite{width}.

 In the SM, at tree level, $\etatautau$ goes through the exchange of a $Z$-boson and we obtain a tiny branching ratio 
$\mathcal{BR}^{Z}( \eta_b\rightarrow \tau^+\tau^-)=3.8\times 10^{-9}$. In our calculation we have used $\Gamma_{Z}=2.4952\pm 0.0023$
 GeV \cite{pdg}.
 {} For the two photon contribution to $\etatautau$, we obtain, using Eq.~\ref{etahigh} and Eq.~\ref{etagamma}, 
$ \mathcal{BR}^{2\gamma}[\etatautau] \ge 4.6 \times 10^{-10}$ for $m_b=4.8$ GeV. Using Eq.~\ref{gammatotal} the total branching ratio for
$\etatautau$ is $\approx 4.3 \times 10^{-9}$.

\begin{figure}[htb!]
\centering
\includegraphics[width=7cm]{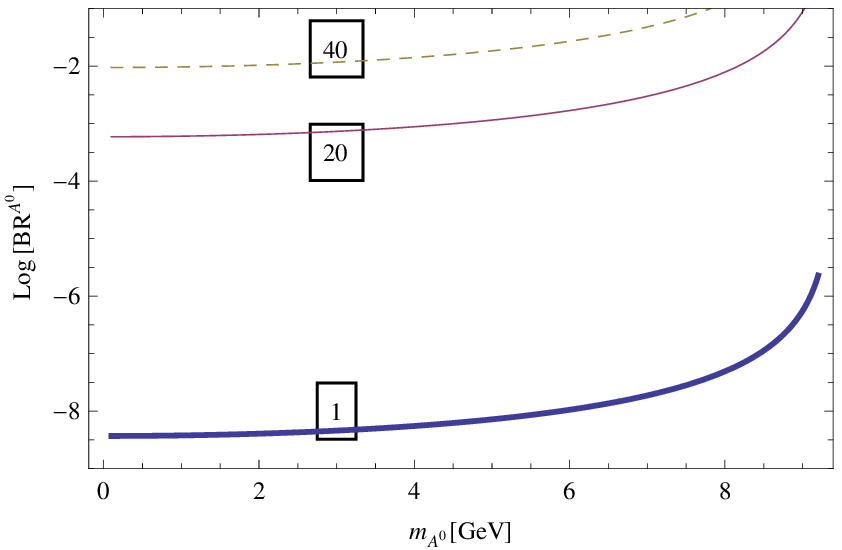}
\includegraphics[width=7cm]{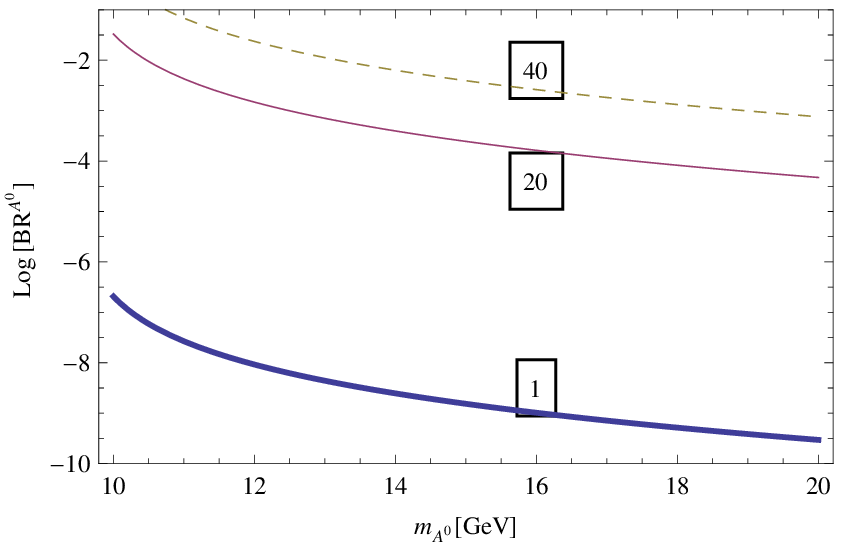}
\caption{The logarithm of $\mathcal{BR}^{A^0}( \eta_b\rightarrow \tau^+\tau^-)$ as a function of $m_{A^{0}}$ for 
different values of $F_{A^{0}}$ and $m_{A^{0}}\in [0.1, 20]$ GeV.}
\label{2HDM}
\end{figure} 
 In Fig.~\ref{2HDM}, we plot the logarithm of the branching ratio for $\etatautau$ mediated by the pseudoscalar $A^0$ in a generic 2HDM model. The branching ratio, $\mathcal{BR}^{A^0}$, is plotted for various values of the $A^{0}$ mass, which we take from 0.1 to 20 GeV, and for various values of $F_{A^{0}}$.  As the mass of the $A^0$ approaches the mass of the $\eta_b$ the branching ratio increases and blows up at
 $m_{A^0}=m_{\eta_b}$. This behavior clearly does not represent the physical situation because in this region the width of the $A^0$ and mixing effects of the $A^0$ with $\eta_b$ become important and regularize the $A^0$ contribution.  We  observe in Fig.~\ref{2HDM} that the branching ratio $ \sim F_{A^0}^4$ is very sensitive to
 $F_{A^0}$.
  The branching ratio is relatively less sensitive to the mass $m_A^0$.   We see from the plots in Fig.~\ref{2HDM} that the branching ratio for $\etatautau$, through the $A^0$ exchange, can be considerably larger than the SM branching  ratios and
  can vary from $ \sim 10^{-8}$ to the experimental bound of 8 \% for $F_{A^{0}}=40$. 
  Since we have neglected the width and mixing effects our predictions are no longer reliable as the mass of the $A^0$ approaches the mass of the $\eta_b$.
  The mixing effects are model dependent
  and as an example, for the model for mixing employed in Ref.~\cite{mix},
  the effects of mixing are important
  in the  $m_{A^0}$ mass range of $9.4 - 10.5$ GeV.
  We see from Fig.~\ref{2HDM} that even outside this range the
  branching ratio for $\etatautau$ can be significant and we expect the same to be true also in the mass range where mixing effects are important.
\begin{figure}[htb!]
\centering
\includegraphics[width=7cm]{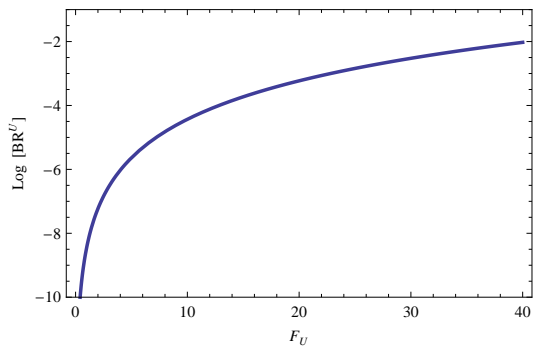}
\caption{The logarithm of $\mathcal{BR}^{U}( \eta_b\rightarrow \tau^+\tau^-)$ as a function of $F_{U}$. }
\label{U-Boson2}
\end{figure}

\begin{figure}[htb!]
\centering
\includegraphics[width=7cm]{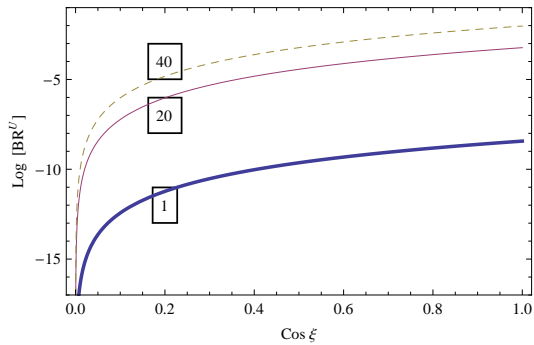}
\caption{The logarithm of $\mathcal{BR}^{U}( \eta_b\rightarrow \tau^+\tau^-)$ as a function of $\cos \zeta$ for different values of $\tan \beta$ and $\cos \zeta \in [0, 1]$. }
\label{U-Boson}
\end{figure}

{}  As discussed in the previous section, the branching ratio for  the decay 
$\mathcal{BR}^{U}( \eta_b\rightarrow \tau^+\tau^-)$ is independent of the mass of the gauge 
boson $U$ in the approximation of neglecting the width of the $U$-boson. We next plot in Fig. \ref{U-Boson2}
 the logarithm of the branching ratio for $\etatautau$   versus $F_{U}$.
 Working in a specific model \cite{Bouchiat, fayet2,fayet3}
  $F_{U}\equiv \cos{\zeta} \tan\beta$. We plot the branching ratio versus the invisibility factor $\cos \zeta$ for different values of $\tan \beta$ in Fig. \ref{U-Boson}.
Again we observe that the branching ratio can vary over a wide range and can be much larger than the SM prediction.

\section{Conclusion} In this paper we explored the decay $\etatautau$ as a probe for a light pseudoscalar or a light axial vector state.
We estimated the SM branching ratios for $\etatautau$ via the $Z$ exchange and the two photon intermediate state and found it to be very small $ \sim 4 \times 10^{-9}$.
We then considered the decay process $\eta_{b}\rightarrow \tau^{+}\tau^{-}$ mediated via the pseudoscalar Higgs boson $A^{0}$ in a 2HDM type NP model. We found that the branching ratio for $\etatautau$ can be substantially larger than the SM prediction and can reach the experimental bound of 8 \%.
Working in a specific model containing a light axial vector state, $U$,
 a similar result was obtained for the branching ratio of $\etatautau$. We also obtained an interesting
  result  that the $\mathcal{BR}^{U}(\etatautau)$ 
is independent of the mass of $U$-boson if the width of the $U$ is neglected. This result followed from the fact that the axial $U$-boson couplings to fermions were  proportional to the mass $m_{U}$ and the fact that $\eta_b$ is a pseudoscalar. A constraint on the $U$-boson mass 
could be obtained by requiring its coupling to fermions to be $\le 1$. In light of the results obtained in the paper an experimental measurement of the branching ratio for $\etatautau$ is strongly desirable as this measurement might reveal the presence of light, $ \sim$ GeV, pseudoscalar or axial vector states. The experimental measurements of $\etatautau$ may be feasible at planned high luminosity B factories and at hadron colliders such as the Tevatron and the LHC,
specially if the branching ratios are much larger than the SM rate.

\end{document}